# Raman Spectra and Strain Effects in Bismuth Oxychalcogenides

Ting Cheng[1,2], Congwei Tan[1,2], Shuqing Zhang[3], Teng Tu[1]

Hailin Peng[1,2,4,*], and Zhirong Liu[1,2,4,*]

[1] College of Chemistry and Molecular Engineering, Peking University, Beijing 100871, China.
[2] Center for Nanochemistry, Academy for Advanced Interdisciplinary Studies, Peking University, Beijing 100871, China
[3] The Low-Dimensional Materials and Devices Laboratory, Tsinghua-Berkeley Shenzhen Institute, Tsinghua University, Shenzhen 518055, Guangdong, China
[4] State Key Laboratory for Structural Chemistry of Unstable and Stable Species, Beijing National Laboratory for Molecular Sciences, Peking University, Beijing 100871, China
[*] Address correspondence to hlpeng@pku.edu.cn and LiuZhiRong@pku.edu.cn

## Abstract

A new type of two-dimensional layered semiconductor with weak electrostatic but not van der Waals interlayer interactions, $Bi_2O_2Se$, has been recently synthesized, which shown excellent air stability and ultrahigh carrier mobility. Herein, we combined theoretical and experimental approaches to study the Raman spectra of $Bi_2O_2Se$ and related bismuth oxychalcogenides ($Bi_2O_2Te$ and $Bi_2O_2S$). The experimental peaks lie at 160 cm$^{-1}$ in $Bi_2O_2Se$ and at 147 cm$^{-1}$ and 340 cm$^{-1}$ in $Bi_2O_2Te$. They were fully consistent with the calculated results (159.89 cm$^{-1}$, 147.48cm$^{-1}$ and 340.33 cm$^{-1}$), and were assigned to the out-of-plane $A_{1g}$, $A_{1g}$ and $B_{1g}$ modes, respectively. $Bi_2O_2S$ was predicted to have more Raman-active modes due to its lower symmetry. The shift of the predicted frequencies of Raman active modes was also found to get softened as the interlayer interaction decreases from bulk to monolayer $Bi_2O_2Se$ and $Bi_2O_2Te$. To reveal the strain effects on the Raman shifts, a universal theoretical equation was established based on the symmetry of $Bi_2O_2Se$ and $Bi_2O_2Te$. It was predicted that the doubly degenerate modes split under in-plane uniaxial/shear strains. Under a rotated uniaxial strain, the changes of Raman shifts are anisotropic for degenerate modes although $Bi_2O_2Se$ and $Bi_2O_2Te$ were usually regarded as isotropic systems similar to graphene. This implies a novel method to identify the crystallographic orientation from Raman spectra under strain. These results have important consequences for the incorporation of 2D Bismuth oxychalcogenides into nanoelectronic devices.

## Introduction

With the boom of studies on two-dimensional (2D) materials, some "star materials" such as graphene,[1-2] transition-metal dichalcogenide[3] and black phosphorene[4-5] have attracted great interest due to their fascinating properties and potential applications. They can all be categorized as van der Waals (vdWs) materials considering the existence of vdWs interaction between layers in their parent counterparts. Very recently, a new type of 2D material with weak interlayer electrostatic interactions instead of vdWs interaction, $Bi_2O_2Se$, has been successfully synthesized.[6] $Bi_2O_2Se$ was found to possess both a suitable bandgap and an ultrahigh Hall mobility at low temperatures, and more importantly, to exhibit excellent environmental stability even after exposure to air for months.[6] Actually, $Bi_2O_2Se$ was initially studied in its bulk (ceramics) form as potential *n*-type thermoelectric material.[7-10] After $Bi_2O_2Se$ was synthesized into ultrathin crystal films with a thickness between tens to two layers,[6, 11] it was recognized as a superior semiconductor, e.g., field-effect transistors made from it exhibit a current on/off ratio larger than $10^6$ at room temperature.[6] There exists a strong spin–orbit interaction in 2D $Bi_2O_2Se$ sheets,[12] and they were shown to have excellent optoelectronic performance as photodetectors.[13-14] The ultrathin thickness may also greatly enhance thermoelectric properties.[15] Inspired by the synthesis of $Bi_2O_2Se$ sheets, attentions were also paid to the related class of bismuth oxychalcogenides such as $Bi_2O_2Te$ and $Bi_2O_2S$,[16-17] which were predicted to possess intriguing piezoelectricity and ferroelectricity.[17] They have adjustable bandgaps in a range between 0.2 eV and 1.5 eV,[17-19] and have little mismatched lattice parameters, yielding their outstanding contributions to the 2D material family.

For the studies of 2D materials, Raman spectroscopy is a valuable tool in analyzing structure, defect and phase transition.[20-22] Incorporated with experimental measurements, theoretical calculations can provide additional microscopic insights which are essential for interpreting and utilizing the Raman spectroscopic information. In the previous works on the $AT_2X_2$ ($A$= K, Rb, Cs, Tl; $T$= Fe, Co, Ni and $X$= S, Se, Te) family which adopt a tetragonal layered crystal structure (space group $I4/mmm$, No.139) similar to $Bi_2O_2Se$ and $Bi_2O_2Te$, the combination of experimental and theoretical Raman spectra shows remarkable advantages.[23-25] So far, detailed theoretical analyses on the Raman properties of the bismuth oxychalcogenides are still lacking.

This work is also motivated by the importance of strain effect. The presence of strain in low-dimensional materials modifies the crystal phonons, band structure and carrier mobility.[26-31] The strain effects are important in understanding the performance of nanoelectronic devices, particularly in the flexible electronics. Usually, tensile strains result in mode softening, while compressive strains result in mode hardening. The shift of phonon frequencies with hydrostatic strain could also give the Grüneisen parameter, which is closely related with thermomechanical properties.[32] In addition, combining the applied strain and Raman shifts, one may identify the crystallographic orientation as previously shown in graphene, $MoS_2$ and $WS_2$.[27, 31, 33]

In this work, we studied the Raman spectra of $Bi_2O_2Se$, $Bi_2O_2Te$ and $Bi_2O_2S$ as well as the corresponding strain effects. We firstly briefly discussed the geometric and electronic structures of these materials, and then systematically studied the first-order Raman active modes using the group factor analysis combined with first-principle calculations and unpolarized Raman scattering experiments. To further compared with the experiments, the experimental geometry configuration and polarization Raman scattering were also discussed here. Lastly, we explored the Raman shifts in the bulk and monolayer $Bi_2O_2Se$ and $Bi_2O_2Te$ with applied mechanical in-plane uniaxial/shear strains. Rotational uniaxial strain was also considered to reveal the anisotropic effect and provide a novel potential method to identify the crystallographic orientation of samples by means of the Raman shifts of split degenerate Raman modes.

# 1. Theoretical and experimental methods

## 1.1 First-principle calculations

The calculations of the geometric optimization and electronic structure were carried out within density functional theory (DFT), and the phonon frequencies were obtained within the density functional perturbation theory (DFPT) method implemented in QUANTUM ESPRESSO package.[34] To obtain the phonon frequencies, we adopted the norm-conserving pseudopotentials with the local density approximation (LDA) exchange correlation function which is more suitable to phonon calculation.[35] However, LDA usually underestimates the bandgap. Therefore, to better verify the electron density using in the DFPT method, we calculated the band dispersion under LDA, and compared the trend with the results using the Perdew-Burke-Ernzerhof-modified Becke-Johnson (PBE-MBJ) exchange-correlation function which is more suitable to obtain an accurate gap.[36] For the reason of computational efficiency, we performed the MBJ calculation implemented in Vienna ab initio simulation package (VASP).[37-38] Spin-orbit coupling (SOC) effects were also considered in both LDA and PBE-MBJ calculations. In the self-consistent field calculation, the kinetic and charge density energy cutoffs were set to 80 Ry and 320 Ry, respectively, when using the Quantum Espresso package, whereas the cutoff kinetic energy was set to 520 eV when using the VASP. The Brillouin zones were sampled with a 24×24×24 and 8×8×1 Monkhorst-Pack **k**-space mesh for the bulk primitive cell and monolayer $Bi_2O_2Se$ and $Bi_2O_2Te$, respectively, and with a 27×27×9 mesh for bulk $Bi_2O_2S$. The convergence criterion for the total energy was set to $10^{-9}$ Ry and atomic forces were smaller than $10^{-5}$ Ry/a.u. for the phonon calculations. For monolayer calculations, interactions between the adjacent layers were limited by setting vacuum intervals of at least 15 Å. All the Raman spectra shown are plotted after uniform Gaussian broadening.

## 1.2 Material synthesis and spectra measurements

The bulk $Bi_2O_2Se$ and $Bi_2O_2Te$ were synthesized using an oxygenation method. In this case, the $O_2$ was directly introduced to induce a phase transformation from $Bi_2Se_3$ to $Bi_2O_2Se$ and from $Bi_2Te_3$ to $Bi_2O_2Te$ (Figure S1). Herein, bulk $Bi_2Se_3$ and $Bi_2Te_3$ were employed and the trace amount of $O_2$ was controlled by using a gas flow meter. The pressure of the system was kept at 200 Torr and the synthesis temperature range was about 590-620°C. Raman spectra were collected with a confocal Raman spectrometer (Jobin Yvon LabRAM HR800) using a He-Ne laser (632.8 nm).

## 2. Results and Discussion

## 2.1 Geometric and electronic structures

We investigated three bismuth oxychalcogenide materials: $Bi_2O_2Se$, $Bi_2O_2Te$ and $Bi_2O_2S$. Their optimized bulk crystal structures are shown in Figure1. They all possess the inversion symmetry. The typical structures consist of stacked BiO layers, sandwiched by X (X=Se, Te and S) ions arrays with relatively weak electrostatic interactions. In each BiO layer, one O atom is bonded to four Bi atoms to form an $OBi_4$ tetrahedral.

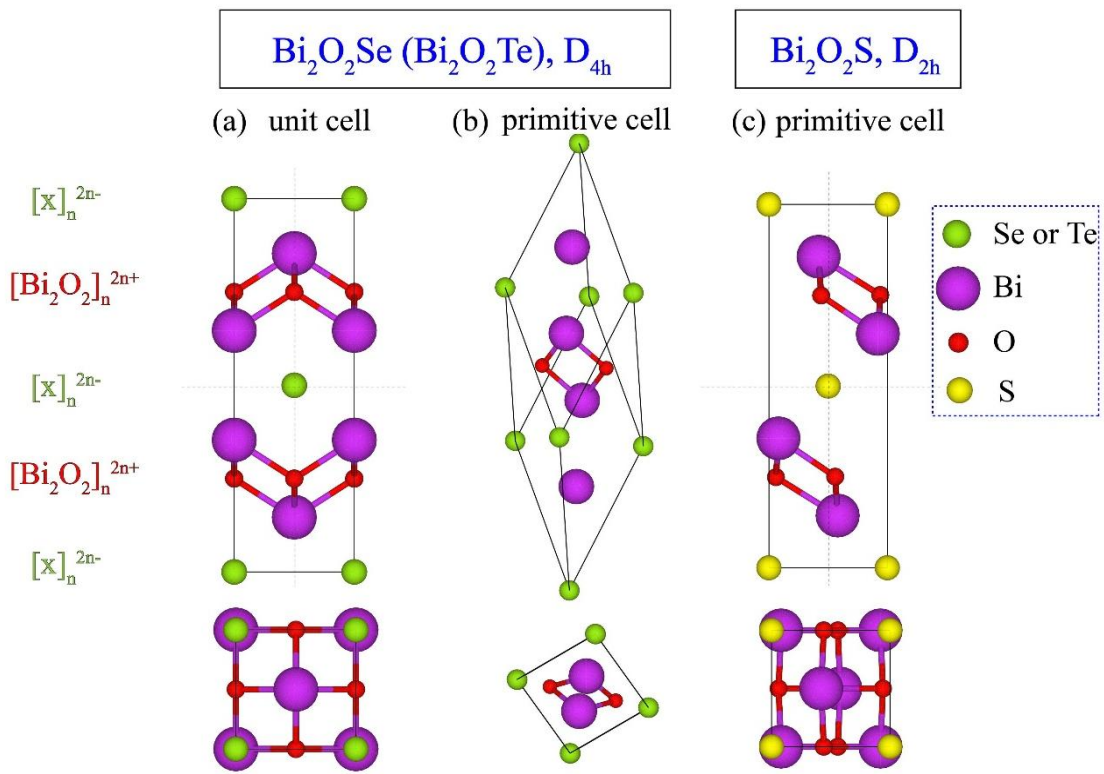

**Figure 1.** Geometric structures of (a) the unit cell of $Bi_2O_2Se$ ($Bi_2O_2Te$), (b) the primitive cell of $Bi_2O_2Se$ ($Bi_2O_2Te$), and (c) $Bi_2O_2S$. The dotted lines in the side view are for the convenience to mark the central axes.

$Bi_2O_2Se$ and $Bi_2O_2Te$ were both experimentally observed to be tetrahedral with space group *I*4/*mmm*,[18, 39] which have ten atoms in the unit cell. Although ferroelectric phases with lower symmetry were predicted elsewhere,[17] they have not been observed in experiments yet. Here we adopted the *I*4/*mmm* symmetry as a constraint in our calculations so that the resulting Raman spectra below can be used to interpret experimental results. As seen from Figure 1 (a), the Se/Te atoms locate at the vertexes and the center of the cell, Bi atoms locate at the vertical edges and the vertical axis of the cell, while the O atoms locate at vertical middle lines of the side faces. In other words, all atoms locate at high-symmetry points or lines. However, the tetrahedral cell is not the smallest repeating unit. Actually, it is composed with two formula units. The smallest primitive unit cell is rhombohedral and contains five atoms [Figure 1 (b)], which will be used to perform the Raman calculation. Different from the highly symmetrical $Bi_2O_2Se$ and $Bi_2O_2Te$, the primitive cell of $Bi_2O_2S$ is not rhombohedral but orthorhombic (with space group *Pnnm*) with anisotropic lattice parameters ($a \neq b$ in the *x*-*y* plane) as revealed in the experiment.[40] S atoms still locate at the vertexes and the center of the unit cell, but Bi and O atoms all deviate from the high-symmetry lines [Figure 1 (c)]. More specifically, the upper BiO layer shifts to the left (minus *x* axis) with a displacement of 0.08 Å for Bi atoms and 0.059 Å for O atoms (in our relaxed structure using LDA method), while the lower BiO layer shifts to the right with the same displacements. Therefore, $Bi_2O_2S$ has a slightly distorted structure. The calculated structural parameters (both in LDA and PBE method) agree to better than 2% with the experimental values,[18, 39-40] which are summarized in Table 1. The lattice parameters of three systems are close, making it possible to fabricate heterostructures with little lattice mismatch.

**Table 1. The optimized lattice constants for Bi$_2$O$_2$Se, Bi$_2$O$_2$Te and Bi$_2$O$_2$S calculated using LDA and GGA methods.**

| Family | Atom | Space group | a (Å) | | | b (Å) | | | c (Å) | | |
|---|---|---|---|---|---|---|---|---|---|---|---|
| | | | LDA | PBE | Exp.[a] | LDA | PBE | Exp.[a] | LDA | PBE | Exp.[a] |
| Bi$_2$O$_2$S | 10 | *Pnnm* | 3.77 | 3.90 | 3.87 | 3.74 | 3.87 | 3.84 | 11.73 | 12.05 | 11.92 |
| Bi$_2$O$_2$Se | 5 | *I4/mmm* | 3.83 | 3.92 | 3.88 | 3.83 | 3.92 | 3.88 | 12.02 | 12.38 | 12.16 |
| Bi$_2$O$_2$Te | 5 | *I4/mmm* | 3.93 | 4.01 | 3.98 | 3.93 | 4.01 | 3.98 | 12.54 | 12.88 | 12.70 |

[a] The experimental values of Bi$_2$O$_2$S, Bi$_2$O$_2$Se and Bi$_2$O$_2$Te were adopted from reference [40], [39] and [18], respectively, to provide a comparision.

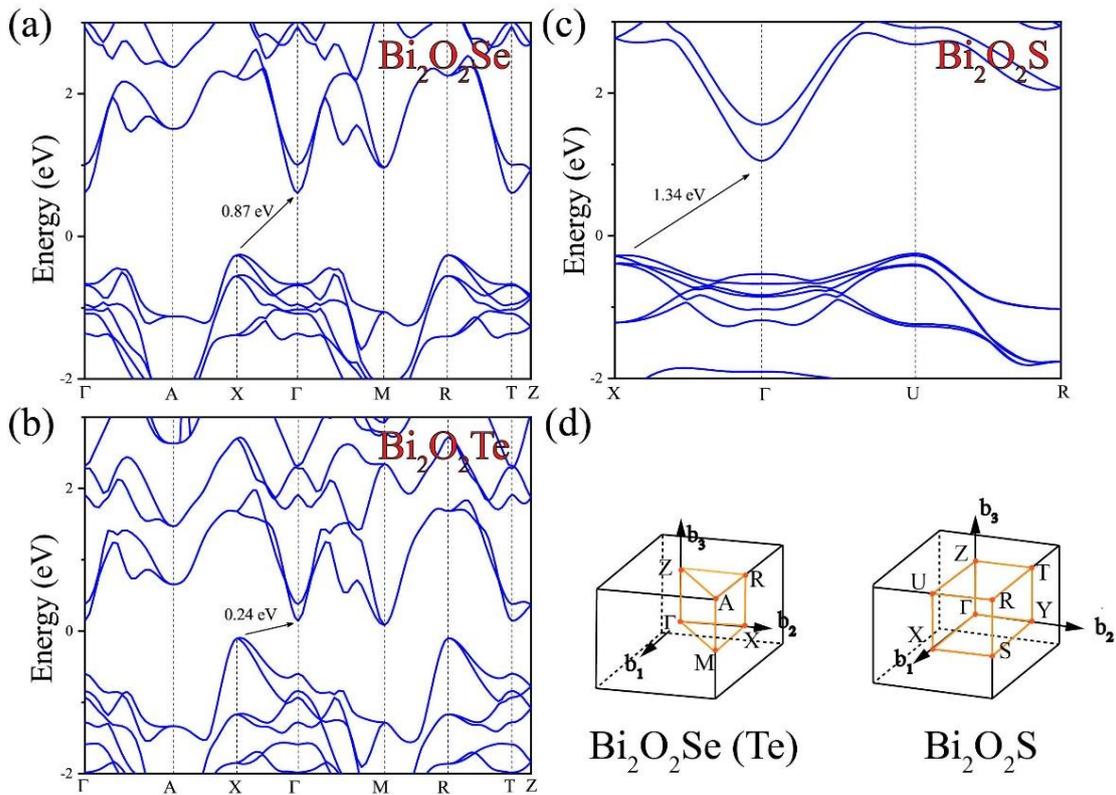

**Figure 2.** Electronic band structures calculated by MBJ method of (a) Bi$_2$O$_2$Se, (b) Bi$_2$O$_2$Te, (c) Bi$_2$O$_2$S and (d) their corresponding first Brillouin zone (following the notation rules in ref. 41). SOC effect was considered here.

The computed band dispersion by MBJ method was displayed in Figure 2. It indicates that three systems are all semiconductors with indirect bandgaps of 0.87, 0.24 and 1.34 eV, respectively, agreeing well with the recent reported experiments on $Bi_2O_2Se$ (0.8 eV)[6] and $Bi_2O_2S$ (1.5 eV)[42] and previous theoretical calculations[17] (0.89, 0.16 and 1.25eV, respectively). Compared with the MBJ method, the LDA method gives similar band dispersion, although the gaps are obvious underestimated as usual (Figure S2).

## 2.2 Raman spectra

### 2.2.1 Factor group analysis and first-principle calculations

Factor group analysis on the *I4/mmm* ($Bi_2O_2Se$, $Bi_2O_2Te$) and *Pnnm* ($Bi_2O_2S$) space group yields a normal mode distribution and Raman tensors at the Brillouin zone center, which are collected in Table 2. For bulk $Bi_2O_2Se$ and $Bi_2O_2Te$ with $D_{4h}$ point group symmetry, five atoms in the primitive cell results in 15 vibrational modes. Γ-point phonons could be represented by the irreducible representations of $D_{4h}$ as: $\Gamma = A_{1g} + 3A_{2u} + B_{1g} + 2E_g + 3E_u$, where one $A_{2u}$ and one $E_u$ are acoustic modes, while $A_{1g}$, $B_{1g}$ and $E_g$ are Raman active. The other $A_{2u}$ and $E_u$ are infrared active. Here the letter "*E*" denotes the doubly degenerate modes in the *xy* plane. For $Bi_2O_2S$ crystals, due to the lower symmetry ($D_{2h}$ point group) and more atoms (10 instead of 5) in the primitive cell, there are more vibrational modes at the Γ point: $\Gamma = 4A_g + 3A_u + 4B_{1g} + 3B_{1u} + 2B_{2g} + 6B_{2u} + 2B_{3g} + 6B_{3u}$, where one $B_{1u}$, one $B_{2u}$ and one $B_{3u}$ are acoustic modes, while $A_g$, $B_{1g}$, $B_{2g}$ and $B_{3g}$ are Raman modes.

The typical vibrational modes of Raman-active phonons and the predicted Raman scattering spectra of the $Bi_2O_2Se$, $Bi_2O_2Te$ and $Bi_2O_2S$ are illustrated in Figure 3. Because $Bi_2O_2Se$ and $Bi_2O_2Te$ share the similar structure with *I4/mmm* group symmetry, they exhibit similar vibrational modes at the Γ-point and four modes ($A_{1g}$, $B_{1g}$, $E_g^1$ and $E_g^2$) among them are Raman active. The atomic displacements associated with the Raman active modes in the primitive and unit cells are both illustrated in Figure 3 (a). For the $A_{1g}$ and $B_{1g}$ mode, they involve the motions of Bi and O atoms along the crystallographic $z$-axis, respectively. On the other hand, the vibrations of Bi and O atoms within the $xy$-plane give rise to two groups of degenerate $E_g$ modes. Therefore, we may identify the $A_{1g}$ and $B_{1g}$ phonons as the breathing modes whereas the two $E_g$

**Table 2. The types of atoms together with their Wyckoff positions and each site's irreducible representations in the Γ-point phonons, as well as Raman tensors, phonon activites and selection rules for Bi$_2$O$_2$Se (Bi$_2$O$_2$Te) and Bi$_2$O$_2$S.**

| | Atoms | Wyckoff position | Irreducible representations |
|---|---|---|---|
| | Bi | 4e (0, 0, z) | $A_{1g} + A_{2u} + E_g + E_u$ |
| | O | 4d (0, 1/2, 1/4) | $B_{1g} + A_{2u} + E_g + E_u$ |
| | Se/ Te | 2a (0, 0, 0) | $A_{2u} + E_u$ |

Bi$_2$O$_2$Se (Te)

Raman Tensors

$$\hat{R}_{A_{1g}} = \begin{pmatrix} a & 0 & 0 \\ 0 & a & 0 \\ 0 & 0 & b \end{pmatrix} \quad \hat{R}_{B_{1g}} = \begin{pmatrix} c & 0 & 0 \\ 0 & -c & 0 \\ 0 & 0 & 0 \end{pmatrix} \quad \hat{R}_{E_g^1} = \begin{pmatrix} 0 & 0 & e \\ 0 & 0 & 0 \\ e & 0 & 0 \end{pmatrix} \quad \hat{R}_{E_g^2} = \begin{pmatrix} 0 & 0 & 0 \\ 0 & 0 & f \\ 0 & f & 0 \end{pmatrix}$$

Activity and selection rules

$$\Gamma_{\text{Raman-active}} = A_{1g}(\alpha_{xx+yy}, \alpha_{zz}) + 2E_g(\alpha_{xz}, \alpha_{yz}) + B_{1g}(\alpha_{xx} - \alpha_{yy})$$

$$\Gamma_{\text{acoustic}} = A_{2u} + E_u$$

| | Atoms | Wyckoff position | Irreducible representations |
|---|---|---|---|
| | Bi, O | 4g (x, y, 0) | $2A_g + A_{2u} + 2B_{1g} + B_{2g} + B_{1u} + 2B_{2u} + B_{3g}$ |
| | S | 2a (0, 0, 0) | $B_{1u} + A_u + 2B_{2u} + 2B_{3u}$ |

Bi$_2$O$_2$S

Raman Tensors

$$\hat{R}_{A_g} = \begin{pmatrix} a & 0 & 0 \\ 0 & b & 0 \\ 0 & 0 & c \end{pmatrix} \quad \hat{R}_{B_{1g}} = \begin{pmatrix} 0 & d & 0 \\ d & 0 & 0 \\ 0 & 0 & 0 \end{pmatrix} \quad \hat{R}_{B_{2g}} = \begin{pmatrix} 0 & 0 & e \\ 0 & 0 & 0 \\ e & 0 & 0 \end{pmatrix} \quad \hat{R}_{B_{3g}} = \begin{pmatrix} 0 & 0 & 0 \\ 0 & 0 & f \\ 0 & f & 0 \end{pmatrix}$$

Activity and selection rules

$$\Gamma_{\text{Raman-active}} = 4A_g(\alpha_{xx}, \alpha_{yy}, \alpha_{zz}) + 4B_{1g}(\alpha_{xy}) + 2B_{2g}(\alpha_{xz}) + 2B_{3g}(\alpha_{yz})$$

$$\Gamma_{\text{acoustic}} = B_{1u} + B_{2u} + B_{3u}$$

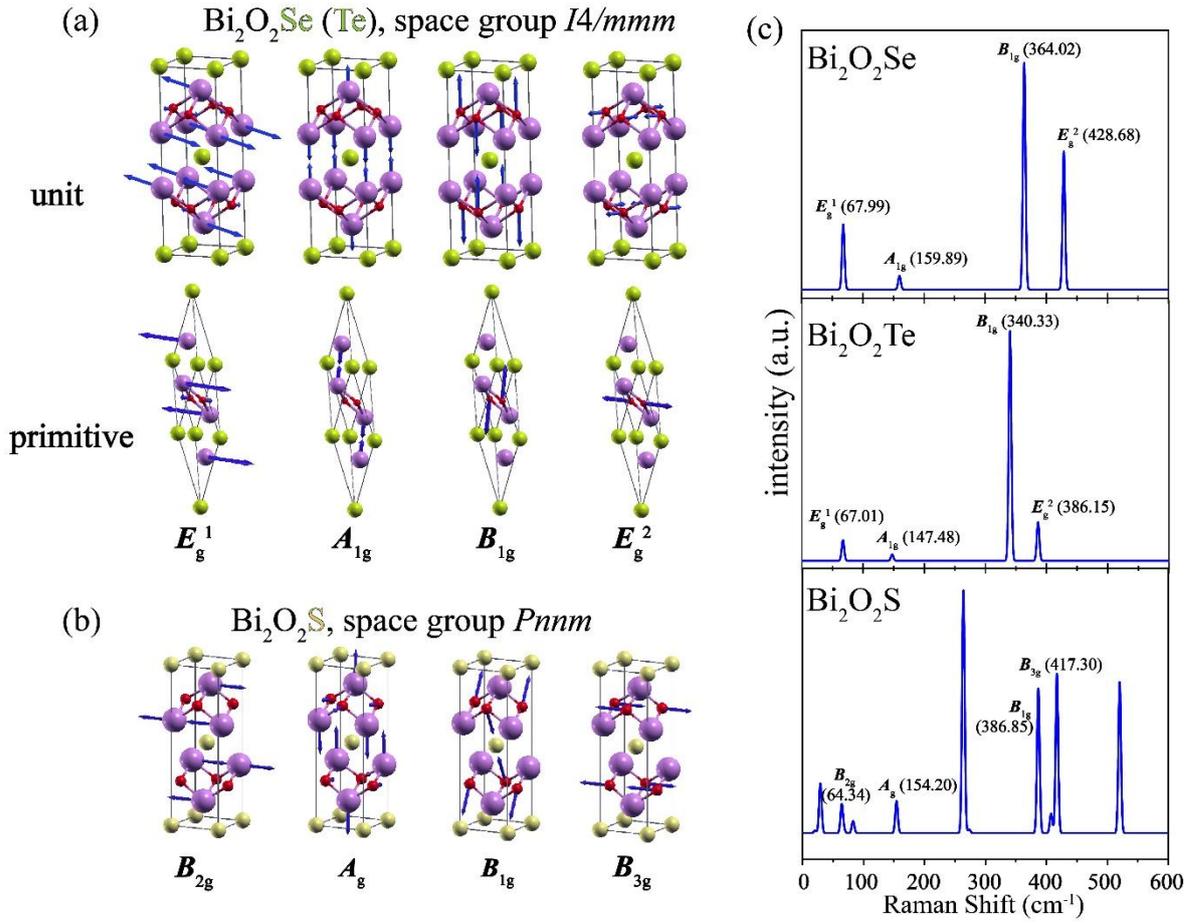

**Figure 3.** Raman spectra of bismuth oxychalcogenides. (a) The atomic displacements of all Raman-active vibrational modes in Bi$_2$O$_2$Se (Bi$_2$O$_2$Te) with a unit cell (top) or a primitive cell (bottom). (b) Four typical Raman-active modes in Bi$_2$O$_2$S, which are similar to those in Bi$_2$O$_2$Se (Bi$_2$O$_2$Te), while the other Raman-active modes can be found in Figure S3. (c) The predicted Raman spectra of Bi$_2$O$_2$Se, Bi$_2$O$_2$Te and Bi$_2$O$_2$S. The frequencies of presented Raman modes in (a, b) are labelled in (c).

modes as the interlayer shear modes. In the case of Bi$_2$O$_2$S, because of its lower symmetry, it possesses more complicated atomic motions (four $A_g$, four $B_{1g}$, two $B_{2g}$ and two $B_{3g}$) at the Γ-point. We just choose four typical modes among them to shown in Figure 3(b), which are similar with those in Bi$_2$O$_2$Se and Bi$_2$O$_2$Te. The other Raman-active modes can be found in

Figure S3. The predicted Raman spectra are shown in Figure 3 (c). The frequencies of four Raman modes in $Bi_2O_2Se$ are 67.99 ($E_g^1$), 159.89 ($A_{1g}$), 364.02 ($B_{1g}$) and 428.68 cm$^{-1}$ ($E_g^2$), respectively, which are in line with the recently calculated results (72, 165.7, 369.4 and 444 cm$^{-1}$, using the PBE method).[43] The corresponding frequencies in $Bi_2O_2Te$ are a litter smaller than those in $Bi_2O_2Se$, which are 67.01 ($E_g^1$), 147.48 ($A_{1g}$), 340.33 ($B_{1g}$) and 386.15 ($E_g^2$) cm$^{-1}$, respectively. For $Bi_2O_2S$ with lower symmetry, the predicted Raman frequencies are in a sequence of 20.52 ($B_{2g}$), 29.23 ($A_g$), 64.34 ($B_{2g}$), 68.23 ($A_g$), 82.86 ($A_g$), 154.20 ($A_g$), 263.85 ($B_{1g}$), 273.27 ($B_{3g}$), 386.85 ($B_{1g}$), 407.76 ($B_{1g}$), 417.30 ($B_{3g}$) and 520.28 ($B_{1g}$) cm$^{-1}$. It is noted that all Raman-active vibrations come from the BiO layer while there is no contribution from Se, Te or S atoms. Therefore, the Raman frequencies of three systems nearly appear in the same range (< 600 cm$^{-1}$). Besides, the predicted peak intensity is expected to be less accurate than the frequency since the experimental peak intensity is usually affected by many factors, such as the substrate, the energy of the incident light, and so on.

**2.2.2 Experimental confirmations**

In general, whether a Raman mode in crystals could be observed in experiments is determined by its Raman tensor and the experimental configuration. In the usual back-scattering configuration, one can only observe two intrinsic Raman peaks ($A_{1g}$ and $B_{1g}$) for $Bi_2O_2Se$ and $Bi_2O_2Te$ according to the Raman tensors shown in Table 2, while for $Bi_2O_2S$, only $A_g$ and $B_{1g}$ modes (eight peaks) are expected to be observed. Observation of the $E_g$ modes for $Bi_2O_2Se$ ($Bi_2O_2Te$) or $B_{2g}$ and $B_{3g}$ modes for $Bi_2O_2S$ requires measurement in the $xz$ or $yz$ planes of the sample, which would be hard for thin 2D sheets but easy for bulk samples.

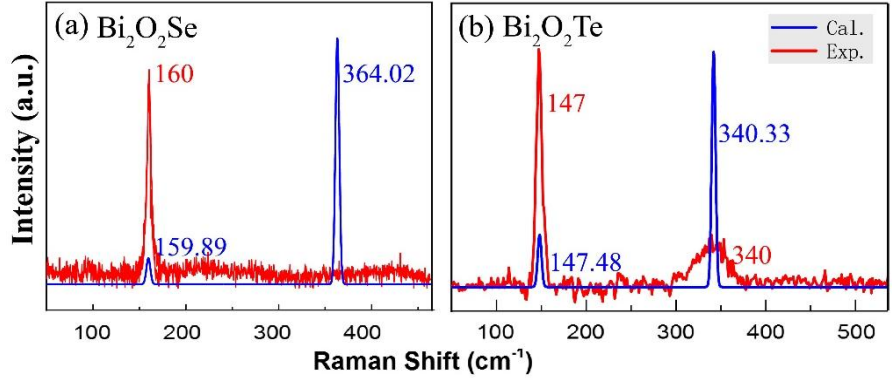

**Figure 4.** Unpolarized Raman scattering spectra of (a) $Bi_2O_2Se$ and (b) $Bi_2O_2Te$ sheet samples (red) measured at room temperature in a back-scattering configuration compared with the theoretical results (blue). The frequencies of peaks are labelled.

A comparison between the experimental results and the predictions is provided in Figure 4 for the unpolarized Raman spectra of $Bi_2O_2Se$ and $Bi_2O_2Te$ in a back-scattering configuration. The experimental peaks located at about 160 cm$^{-1}$ in $Bi_2O_2Se$ and 147 cm$^{-1}$ in $Bi_2O_2Te$ could be assigned as the $A_{1g}$ mode, which are in good match with our predictions (159.89 cm$^{-1}$ and 147.48 cm$^{-1}$, respectively). The experimental peak at the 340 cm$^{-1}$ in $Bi_2O_2Te$ (which is much less sharp) could be assigned as the $B_{1g}$ mode, also being consistent with the prediction (340.83 cm$^{-1}$). In the experiment we did not observe a peak near 364 cm$^{-1}$ (the predicted $B_{1g}$ mode) in $Bi_2O_2Se$. It is unclear whether the $B_{1g}$ peak is absent because the intensity is too weak (noting that the detected intensity of $B_{1g}$ mode is much lower than $A_{1g}$ mode in $Bi_2O_2Te$) or for other unknown reasons. In a recent study of $Bi_2O_2Se$ under high pressure by Pereira *et al.*, the $B_{1g}$ mode was also not observed in the Raman spectrum measurement and only $A_{1g}$ mode (at about 159.2 cm$^{-1}$) existed.[43] They explained the absence of $B_{1g}$ mode in terms of the plasmon-phonon coupling L$^+$ or L$^-$ bands of $B_{1g}$ modes caused by the large carrier concentration in n-type semiconductor.[43] According to previous Raman experiments and calculations on $AT_2X_2$ structures with *I4/mmm* symmetry (such as iron-chalcogenide compounds), defects such as vacancy ordering, consequently, symmetry lowering, may lead to the appearance of additional Raman peaks.[24, 44-45] As an evidence, in a previous high-quality $Bi_2O_2Se$ sample, only one Raman peak located at about 160 cm$^{-1}$ was also reported,[11] where the peak at 160 cm$^{-1}$ can be assigned to the $A_{1g}$ mode In the current case, no additional peaks rather than the intrinsic Raman peaks were observed, indicating the high quality of our synthesized sample. On this occasion, the reason for the disappearing of $B_{1g}$ in $Bi_2O_2Se$ requires further verification of experimental and theoretical studies.

**Table 3. The calculated and experimental Raman frequencies for $Bi_2O_2Se$, $Bi_2O_2Te$, $Bi_2Se_3$ and $Bi_2Te_3$. The frequencies for $Bi_2O_2Te$ and $Bi_2Te_3$ are listed in brackets. The structures were fully or partially relaxed depending on whether the lattice parameters were fixed to the experimental values in calculations or not, resulting in different frequency values as listed.**

|  | $Bi_2O_2Se$ ($Bi_2O_2Te$) space group $I4/mmm$ | | | | $Bi_2Se_3$ ($Bi_2Te_3$) space group $R\bar{3}m$ | | |
|---|---|---|---|---|---|---|---|
|  | Fully relaxed | Partially relaxed | Exp.[a] |  | Fully relaxed | Partially relaxed | Exp.[46-47] |
| $E_g^1$ | 72.21 (72.84) | 67.99 (67.01) | - (-) | $E_g$ | 44.4 (43.2) | 40.4 (40) | 37 (36) |
| $A_{1g}$ | 166.83 (154.25) | 159.89 (147.48) | 160 (147) | $A_{1g}$ | 75 (64.4) | 68 (58.8) | 69 (62) |
| $B_{1g}$ | 376.20 (352) | 364.02 (340.33) | - (340) | $E_g$ | 142.7 (112.8) | 135 (108.9) | 131 (102) |
| $E_g^2$ | 449.60 (408.03) | 428.68 (386.15) | - (-) | $A_{1g}$ | 180.6 (140.8) | 174.2 (138.5) | 174 (133) |

[a] The frequency values of $Bi_2O_2Se$ and $Bi_2O_2Te$ are chosen from our experiments.

A better way to distinguish and assign different Raman modes in experiments is to adopt polarized configurations. According to the selection rules, $A_g$ and $B_{1g}$ modes of $Bi_2O_2S$ could be easily distinguished in parallel and crossed polarization configurations, while more complicated configurations are required to distinguish $A_{1g}$ and $B_{1g}$ modes in $Bi_2O_2Se$ and $Bi_2O_2Te$ (see SI and Figure S4 for more details). Ramon modes in similar systems $AT_2X_2$ (with $I4/mmm$ group symmetry) have been successfully distinguished with such an approach.[24, 48] Limited by the current sample condition, we did not pursue polarized Raman experiments here.

To further verify the reliability of the calculation methods, we performed extra calculations on $Bi_2Se_3$ and $Bi_2Te_3$ to compare with the reported experimental results.[46-47] The Raman frequencies are usually sensitive to the values of lattice constants. Therefore, two schemes of structural optimization were adopted in our calculations. In the first scheme, the structure was fully relaxed where both the lattice constants and the atomic coordinates were optimized. In the second scheme, the structure was partially relaxed, where only the atomic positions were optimized while the lattice parameters were kept at their experimental values. Similar schemes were also applied for $Bi_2O_2Se$ and $Bi_2O_2Te$. The obtained results are summarized in Table 3. It can be seen that the values obtained with fully relaxed structures are obviously larger than the experimental values, owing to the fact that LDA usually underestimates the lattice parameters. In comparison, the results obtained under the partially relaxed structure (with experimental lattice constants) are more consistent with the observed values, with an average deviation as small as 3.5 cm$^{-1}$. Similar trend was also found in other systems such as $MoS_2$.[49] Therefore, to compare with the Raman experiments, it is better to adopt the experimental lattice parameters in calculations. The calculated Raman spectra in Figures 3 and 4 for $Bi_2O_2Se$ and $Bi_2O_2Te$ also adopted such a scheme.

Apart from the bulk materials, we also calculated the Raman frequencies in fully relaxed monolayer $Bi_2O_2Se$ and $Bi_2O_2Te$. The resulting frequencies are 80.44, 153.59, 366.10 and 409.95 cm$^{-1}$ for $Bi_2O_2Se$, and 59.20, 141.10, 340 and 364.35 cm$^{-1}$ for $Bi_2O_2Te$, all exhibiting red shifts from the bulk except the $E_g^1$ mode (80.44 cm$^{-1}$ vs. 72.21 cm$^{-1}$) in $Bi_2O_2Se$. Similar mode softening has also been found in many vdWs layer materials such as graphene and $Bi_2Se_3$, which may be explained by the weaker interlayer restoring forces in the vibrations when the number of layer decreases.[46, 50-52]

## 2.3 Strain dependence of Raman spectra

In this section, we theoretically investigated the Raman shift of monolayer and bulk $Bi_2O_2Se$ and $Bi_2O_2Te$ (all with $D_{4h}$ symmetry) under the mechanical in-plane strain. The applied strain tensor is written as:

$$\boldsymbol{\varepsilon} = \begin{bmatrix} \varepsilon_{xx} & \gamma \\ 0 & \varepsilon_{yy} \end{bmatrix}, \tag{1}$$

where $\varepsilon_{xx}$ and $\varepsilon_{yy}$ represent the uniaxial strain along the $x$ and $y$ directions, respectively, and $\gamma$ is the engineering shear strain which is related to the usual (symmetric) shear strain $\varepsilon_{xy}$ as $\gamma = 2\varepsilon_{xy}$.[30] Here, we focus on the linear effect of strain on vibration frequencies.

**2.3.1 In-plane uniaxial and shear strain**

The Raman frequency $\omega$ is a function of applied strain $\varepsilon$. For a non-degenerate vibrational mode such as $A_{1g}$ and $B_{1g}$, the frequency $\omega$ could be generally expanded as a Taylor series of $\varepsilon$ components:

$$\omega(\varepsilon_{xx}, \varepsilon_{yy}, \gamma) = \omega_0 + k_x \varepsilon_{xx} + k_y \varepsilon_{yy} + k_\gamma \gamma + \text{(high-order terms)}, \tag{2}$$

where $k_x$, $k_y$ and $k_\gamma$ are the linear coefficients, which could be determined by separately applying the strain $\varepsilon_{xx}$, $\varepsilon_{yy}$ and $\gamma$. Under the D$_{4h}$ symmetry of Bi$_2$O$_2$Se and Bi$_2$O$_2$Te, the *x*-axis and *y*-axis are equivalent, which requires that $k_x = k_y \equiv \bar{k}$. Moreover, under a mirror reflection with a normal direction along *x*- or *y*-axis, $\gamma$ changes into $-\gamma$, and the symmetry requires that $k_\gamma = 0$. Therefore, with the D$_{4h}$ symmetry, Eq. (2) is simplified into

$$\omega(\varepsilon_{xx}, \varepsilon_{yy}, \gamma) = \omega_0 + \bar{k}(\varepsilon_{xx} + \varepsilon_{yy}) + \text{(high-order terms)} \tag{3}$$

with only one independent parameter $\bar{k}$ under the linear expansion. When the mode is degenerate, it will split into two bands under strains and the relationship between $\omega$ and $\varepsilon$ should be analyzed by examining the strain response of the corresponding Hamiltonian.[53] Under the D$_{4h}$ symmetry, the splitting of degenerate modes such as $E_g$ is given as (the detailed deduction is given in the SI):

$$\omega_{1,2}(\varepsilon_{xx}, \varepsilon_{yy}, \gamma) = \omega_0 + \bar{k}(\varepsilon_{xx} + \varepsilon_{yy}) \pm \sqrt{[\Delta k(\varepsilon_{xx} - \varepsilon_{yy})]^2 + (k_\gamma \gamma)^2} + \text{(high-order terms)}, \tag{4}$$

where $\bar{k}$, $\Delta k$ and $k_\gamma$ are three independent parameters under the D$_{4h}$ symmetry. It is noted that the number of independent parameters is determined by the symmetry. For example, graphene, graphyne and graphdiyne have a higher symmetry (D$_{6h}$), so $\Delta k = k_\gamma$ and there are only two independent parameters for strain effect of $E_g$ mode.[53]

We conducted first-principle calculations on monolayer and bulk Bi$_2$O$_2$Se and Bi$_2$O$_2$Te with various $\varepsilon_{xx}$, $\varepsilon_{yy}$ and $\gamma$ strains up to $\pm 4\%$. Before discussing the behaviors of the Raman spectrum, we first briefly survey the mechanical properties of the studied systems. The elastic constant $C_{11}$ is critical in determining the mobility.[54-55] The calculated $C_{11}$ of monolayer Bi$_2$O$_2$Se was 100 J m$^{-2}$ (see Figure S5 for the fitted curve), which compares well to the previous calculation with PBE method (92 J m$^{-2}$).[15] For monolayer Bi$_2$O$_2$Te, it was a little smaller, 95 J m$^{-2}$. The shear modulus $C_{66}$ of these two materials are 54 and 50 J m$^{-2}$, respectively. The obtained $C_{11}$ of monolayer Bi$_2$O$_2$Se and Bi$_2$O$_2$Te is only about one-third of that for graphene (358 J m$^{-2}$),[56] and also smaller than that of graphyne (207 Jm$^{-2}$), graphdiyne (159 Jm$^{-2}$) or 8B-*Pmmn* borophene

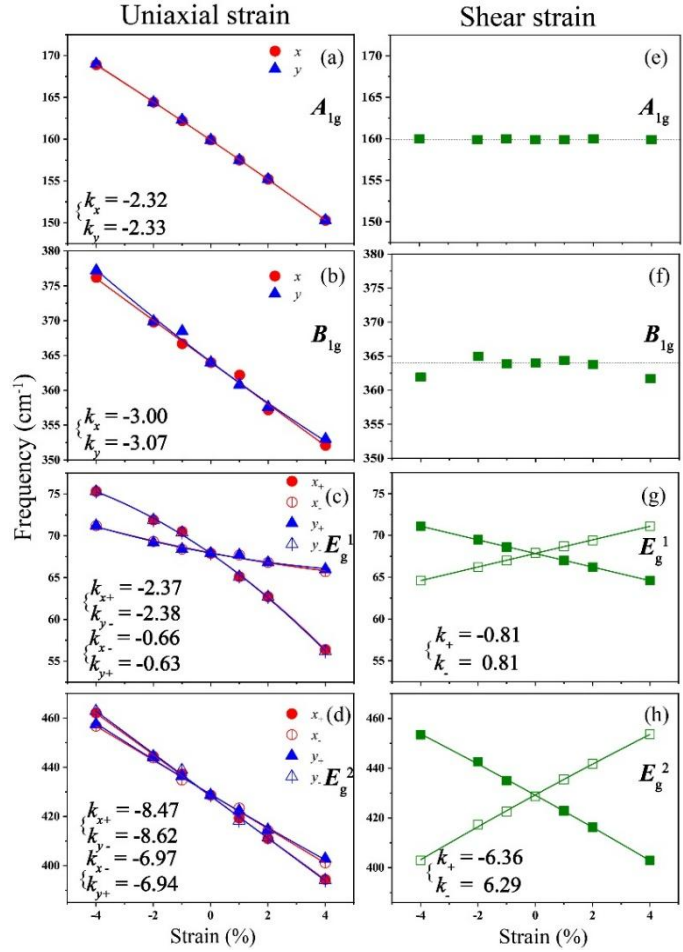

**Figure 5.** The evolutions of Raman shift with (a-d) uniaxial strain and (e-h) shear strain for bulk Bi$_2$O$_2$Se. The calculated data points (squares and triangles) are fitted with a parabolic equation (solid lines) and the fitted linear coefficients (unit: cm$^{-1}$/%) are also shown in the panels. Red and blue are used to denote the results under $\varepsilon_{xx}$ and $\varepsilon_{yy}$, respectively. For the degenerate modes, the open and filled symbols are used to distinguish the two splitting frequencies under the uniaxial strain.

(253 Jm$^{-2}$).[53, 57] But it is comparable or even higher than that of monolayer black phosphorous (29 and 102 Jm$^{-2}$ along the *x* and *y* direction, respectively).[58]

The calculated shift and splitting of Raman frequencies under various strains were displayed in Figure 5 for bulk Bi$_2$O$_2$Se. The data points were fitted with parabolic curves and the resulting linear coefficients (slopes) were listed in the figure. The fitting works well. Under the D$_{4h}$ symmetry, the *x*- and *y*-directions are equivalent. For nondegenerate modes, the slopes with respective to $\varepsilon_{xx}$ and $\varepsilon_{yy}$ were practically identical (−2.32 ≈ −2.33 for $A_{1g}$ mode and −3.00 ≈ −3.07 for $B_{1g}$ mode), being consistent with the analysis in Eq. (3). For each doubly degenerate $E_g$ mode, it splits into two branches and gives four linear coefficients under $\varepsilon_{xx}$ and $\varepsilon_{yy}$ in a direct fitting. They are well grouped into two values (*e.g.*, −2.37 ≈ −2.38 and −0.66 ≈ −0.63 for $E_g^1$ mode), which agrees well with our Eq. (4) and can be combined to give $\bar{k}$ and $\Delta k$. Overall, all Raman modes show redshifts under tensile strains and blueshifts under compressive strains. The $E_g^2$ mode shows the largest slope, indicating that it is the most affected under uniaxial strain. Different from the effect of uniaxial strain where two splitting branches of doubly degenerate $E_g$ modes always shift along the same direction, a shear strain would soften one branch while harden another with opposite slopes [Figure 5 (g, h)] as predicted by Eq. (4). For non-degenerate A$_{1g}$ and B$_{1g}$ modes, they are unaffected by shear strain.

We have systematically investigated the strain effects in Bi$_2$O$_2$Se and Bi$_2$O$_2$Te in both bulk and monolayer forms. Detailed calculated results were shown in Figure S6-11, and the extracted coefficients $\bar{k}$, $\Delta k$ and $k_\gamma$ were summarized in Table 4. The coefficients of Bi$_2$O$_2$Se are close to those of Bi$_2$O$_2$Te no matter in the bulk form or the monolayer form, owing to their similar structures and vibration modes. For the differences between the bulk and the monolayer, the coefficients for $B_{1g}$ mode in monolayer are obviously smaller than the bulk ones, whereas those for the $A_{1g}$ mode keep nearly unchanged.

**Table 4. Determined coefficient $\bar{k}, \Delta k$ and $k_\gamma$ (in units of cm$^{-1}$/%) for four Raman-active modes in bulk and monolayer Bi$_2$O$_2$Se and Bi$_2$O$_2$Te under applied strains.**

|  |  | $E_g^1$ | | | $A_{1g}$ | $B_{1g}$ | $E_g^2$ | | |
|---|---|---|---|---|---|---|---|---|---|
|  |  | $\bar{k}$ | $\Delta k$ | $k_\gamma$ | $\bar{k}$ | $\bar{k}$ | $\bar{k}$ | $\Delta k$ | $k_\gamma$ |
| Bi$_2$O$_2$Se | Bulk | −1.51 | 0.85 | 0.81 | −2.33 | −3.04 | −7.72 | 0.75 | 6.33 |
|  | Mono | −2.15 | 0.43 | 1.86 | −2.31 | −1.65 | −5.97 | 0.56 | 6.59 |
| Bi$_2$O$_2$Te | Bulk | −1.88 | 0.95 | 1.05 | −2.32 | −3.53 | −7.68 | 0.93 | 5.78 |
|  | Mono | −1.61 | 0.41 | 1.56 | −2.32 | −1.67 | −6.07 | 0.31 | 6.38 |

The above extracted coefficients of phonon shift under strains could give the Grüneisen parameter $\gamma_m$, which is a parameter related with thermodynamic properties and have been discussed in isotropic 2D materials such as graphene and 2H-MoS$_2$.[27-28] The Grüneisen parameter for a mode $m$, $\gamma_m$, is defined as:[32]

$$\gamma_m = -\frac{1}{\omega_m}\frac{\partial \omega_m}{\partial \varepsilon},$$

where the strain $\varepsilon = \varepsilon_{xx} + \varepsilon_{yy}$ is the hydrostatic component of the applied uniaxial strain. Therefore, based on Eqs. (3, 4), we have $\gamma_m = -\bar{k}/\omega_m$ for both degenerate and non-degenerate modes. Then, the $\gamma_m$ that we found for each mode were: 2.67, 1.50, 0.45 and 1.46 for $E_g^1$, $A_{1g}$, $B_{1g}$ and $E_g^2$ modes in monolayer Bi$_2$O$_2$Se, respectively, and 2.72, 1.64, 0.49 and 1.67 for monolayer Bi$_2$O$_2$Te. These values are comparable to those for the G ($\gamma_m$= 1.72) and 2D ($\gamma_m$=2.0) peaks of graphene,[27, 59] and are larger than those in monolayer 2H-MoS$_2$ (0.54, 0.65, and 0.21 for $E''$, $E'$, and $A_1'$ modes)[28].

### 2.3.2 In-plane rotated uniaxial strain

When a uniaxial strain with a magnitude $\varepsilon$ is applied along a direction with angle $\theta$ to the x-axis, the stain tensor can be expressed by a rotation transformation as:

$$\boldsymbol{\varepsilon}(\varepsilon,\theta) = \begin{bmatrix} \varepsilon_{xx} & \varepsilon_{xy} \\ \varepsilon_{xy} & \varepsilon_{yy} \end{bmatrix} = \mathbf{R}^T \begin{bmatrix} \varepsilon & 0 \\ 0 & 0 \end{bmatrix} \mathbf{R} = \begin{bmatrix} \varepsilon \cos^2\theta & -\varepsilon \sin\theta \cos\theta \\ -\varepsilon \sin\theta \cos\theta & \varepsilon \sin^2\theta \end{bmatrix}, \quad (5)$$

where **R** is the 2D rotation matrix:

$$\mathbf{R} = \begin{bmatrix} \cos\theta & -\sin\theta \\ \sin\theta & \cos\theta \end{bmatrix}. \quad (6)$$

We could then obtain the variation of Raman frequencies under a rotated uniaxial strain by substituting Eq. (5) into Eqs. (3, 4). Considering the linear effects, for a non-degenerate vibrational mode, Eq. (3) evolves into

$$\omega(\varepsilon,\theta) = \omega_0 + \bar{k}\varepsilon. \quad (7)$$

Therefore, the strain effect on non-degenerate modes such as $A_{1g}$ and $B_{1g}$ is isotropic, i.e., the Raman frequencies do not depend on the direction of strain, which is different from anisotropic systems such as black phosphorus.[29] For degenerate modes such as $E_g^1$ and $E_g^2$, Eq. (4) evolves into

$$\omega_{1,2}(\varepsilon,\theta) = \omega_0 + \{\bar{k} \pm \sqrt{[\Delta k \cos(2\theta)]^2 + [k_\gamma \sin(2\theta)]^2}\}\varepsilon, \quad (8)$$

taking into account $\gamma = 2\varepsilon_{xy}$. Hence, the change rate of the frequency versus strain is angle-dependent:

$$k_{1,2} = \frac{\partial \omega_{1,2}}{\partial \varepsilon} = \bar{k} \pm \sqrt{[\Delta k \cos(2\theta)]^2 + [k_\gamma \sin(2\theta)]^2}. \quad (9)$$

The change rates of Raman frequencies under a rotating strain were demonstrated in Figure 6 for $Bi_2O_2Se$. It was recognized from Eq. (4) that the anisotropy depends on the relative values between $\Delta k$ and $k_\gamma$. For the $E_g^1$ mode, the anisotropy is weak [Figure 6(a)] since its $\Delta k$ and $k_\gamma$, are close (0.85 *vs*. 0.81). In contrast, $\Delta k$ and $k_\gamma$ are markedly different (0.75 *vs*. 6.33), so the resulting anisotropic effect is strong [Figure 6(b)]. The anisotropic effects under a rotated strain were also found in monolayer $Bi_2O_2Se$ and $Bi_2O_2Te$ as shown in the supporting information (Figure S12). It is worth noted that degenerate $E_g$ modes in graphene or graphynes also split under a uniaxial strain, but the therein frequencies are isotropic since they have $\Delta k = k_\gamma$ due to the symmetry.

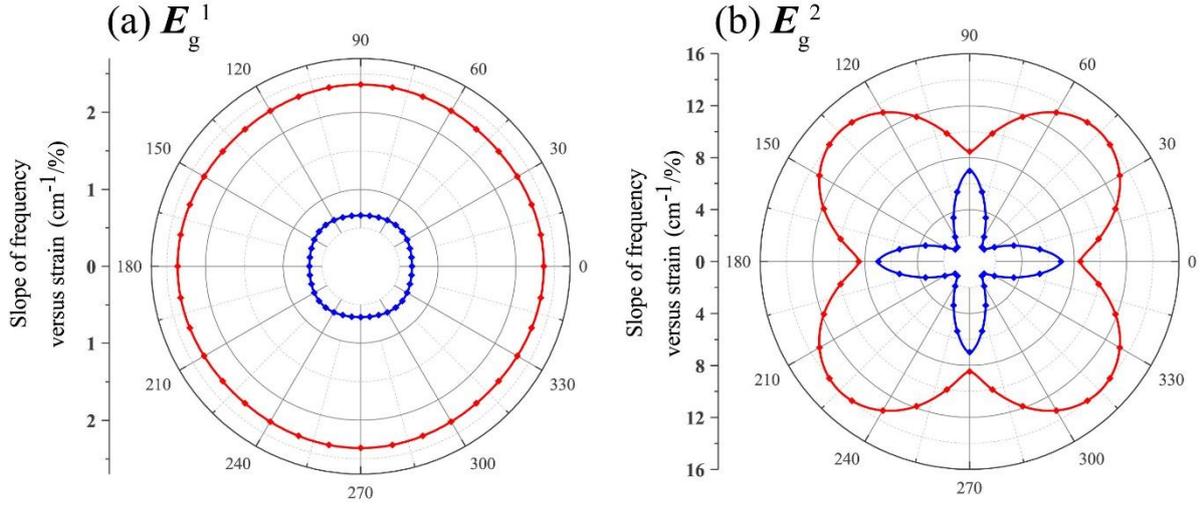

**Figure 6.** The changes of Raman frequencies under a rotated uniaxial strain for degenerate (a) $E_g^1$ and (b) $E_g^2$ modes in bulk $Bi_2O_2Se$. The data were calculated from Eq. (9) with the values of $\bar{k}$, $\Delta k$ and $k_\gamma$ listed in Table 4. Red and blue lines represent $k_1$ and $k_2$ for two splitting bands, respectively.

The anisotropic strain effect can be utilized to determine the crystallographic orientation of $Bi_2O_2Se$ and $Bi_2O_2Te$ in experiments. When an experimental sample is stretched along a fixed direction, $k_{1,2}$ can be readily measured as the ratio between the change of frequency and the magnitude of strain, and then the orientation angle $\theta$ can be solved from Eq. (9) with the known parameters $\bar{k}$, $\Delta k$ and $k_\gamma$ listed in Table 4. A practical scheme is provided here. Denote

$$\begin{cases} a = \cos^2(2\theta) \\ b = \sin^2(2\theta) \end{cases} \quad (10)$$

then a recombination of Eq. (9) gives

$$\begin{bmatrix} (k_1 - k_2)^2 \\ (k_1 + k_2)^2 \end{bmatrix} = \begin{bmatrix} (\Delta k)^2 & (k_\gamma)^2 \\ \bar{k}^2 & \bar{k}^2 \end{bmatrix} \begin{bmatrix} a \\ b \end{bmatrix} \quad (11)$$

After $a$ and $b$ are solved from Eq. (11), $\theta$ is determined to be

$$\theta = \begin{cases} 0° & \text{(if } b < 0\text{)} \\ 45° & \text{(if } a < 0\text{)} \\ \frac{1}{2}\arccos\sqrt{\frac{a}{a+b}} & \text{(otherwise)} \end{cases} \quad (12)$$

The strain efficiency, i.e., the efficiency of a strain to transfer from the substrate to the sample, does not affect the accuracy of the determined $\theta$ as the previous study on black phosphorus.[29]

## 3. Conclusions

In summary, based on group-theory analysis and first-principles calculations, we predicted the Raman spectra of $Bi_2O_2Se$, $Bi_2O_2S$ and $Bi_2O_2Te$, and determined their response to mechanical strain. $Bi_2O_2Se$ and $Bi_2O_2Te$ share the similar structure with $D_{4h}$ point symmetry and four Raman-active modes ($A_{1g}$, $B_{1g}$ and $2E_g$). Experimental work was carried out to give the Raman spectra of $Bi_2O_2Se$ and $Bi_2O_2Te$. The observed peaks (160 cm$^{-1}$ in $Bi_2O_2Se$ and 147 cm$^{-1}$ and 340 cm$^{-1}$ in $Bi_2O_2Te$) were fully consistent with the calculated results (159.89 cm$^{-1}$, 147.48cm$^{-1}$ and 340.33 cm$^{-1}$) and were assigned to the out-of-plane $A_{1g}$, $A_{1g}$ and $B_{1g}$ modes, respectively. $Bi_2O_2S$, on the other hand, possesses a lower symmetry ($D_{2h}$) and more atoms in its unit cell, so it has more Raman-active modes ($4A_g$, $4B_{1g}$, $2B_{2g}$ and $2B_{3g}$). From bulk to the monolayer $Bi_2O_2Se$ and $Bi_2O_2Te$, most Raman frequencies show redshift, which may result from the decrease of the interlayer interactions as that in other 2D materials. To reveal the strain effects on the Raman shifts, a universal theoretical equation was established based on $D_{4h}$ symmetry. All Raman modes show redshifts under the tensile strain. The doubly degenerate modes split under uniaxial and shear strains. Under a rotated uniaxial strain, the frequency variation of degenerate modes is anisotropic although the *x*- and *y*-axes are equivalent under the $D_{4h}$ symmetry, being in contrast to the isotropic case of graphene with $D_{6h}$ symmetry. This provides a novel method to identify the crystallographic orientation of the $D_{4h}$ system by measuring the changes of degenerate Raman modes under an applied uniaxial strains. Overall, the changes of Raman shift under strain and the good match in the lattice constants are beneficial for bismuth oxychalcogenides-based nanoelectronics and mechanical systems.

## Acknowledgements

The authors thank Shiqiao Du, Hengxin Tan, Zuzhang Lin, Prof. Satoshi Watanabe, Prof. Emi Minamitani, Zeyuan Ni and Lin Zhou for helpful discussions. This work was supported by the National Natural Science Foundation of China (Grant Nos. 21773002, 21733001).

# Supporting Information

Ting Cheng[1,2], Congwei Tan[1,2], Shuqing Zhang[3], Teng Tu[1],

Hailin Peng[1,2,4,*], and Zhirong Liu[1,2,4,*]

[1] College of Chemistry and Molecular Engineering, Peking University, Beijing 100871, China.
[2] Center for Nanochemistry, Academy for Advanced Interdisciplinary Studies, Peking University, Beijing 100871, China
[3] The Low-Dimensional Materials and Devices Laboratory, Tsinghua-Berkeley Shenzhen Institute, Tsinghua University, Shenzhen 518055, Guangdong, China
[4] State Key Laboratory for Structural Chemistry of Unstable and Stable Species, Beijing National Laboratory for Molecular Sciences, Peking University, Beijing 100871, China
[*] Address correspondence to hlpeng@pku.edu.cn and LiuZhiRong@pku.edu.cn

1. **SEM images of the sample**

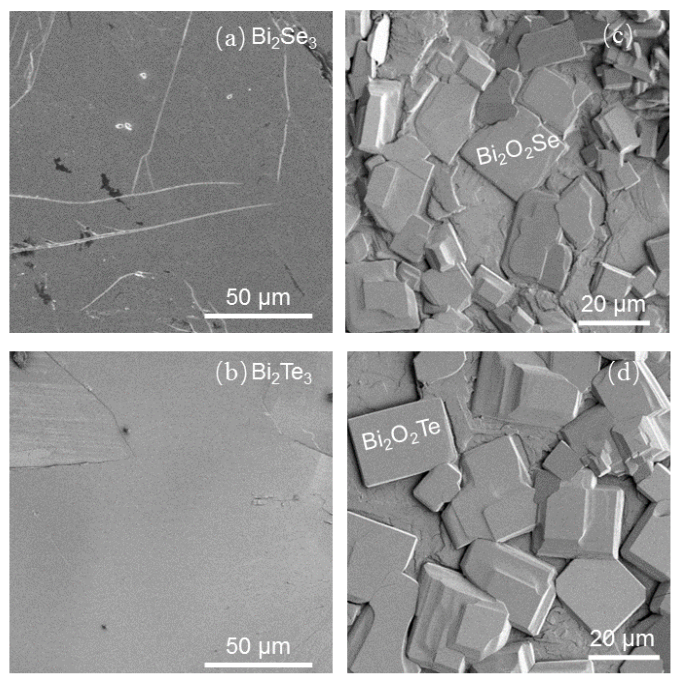

**Figure S1.** SEM images of bulk (a) $Bi_2Se_3$, (b) $Bi_2Te_3$ and CVD-grown (c) $Bi_2O_2Se$, (d) $Bi_2O_2Te$. It can be seen an obvious phase transformation from $Bi_2Se_3$ ($Bi_2Te_3$) to $Bi_2O_2Se$ ($Bi_2O_2Te$).

2. **The calculated Band structure using the LDA method.**

Limited by our server computing performance and low calculation efficiency of MBJ or HSE method in Quantum-Espresso package, we just do the band calculation using the LDA method in

Quantum-Espresso package and to have a qualitative understanding of band dispersion. And also compare the dispersion with the results using the PBE-MBJ method implemented in VASP along the same high-symmetry path in the Brillouin zone to verify the electron density.

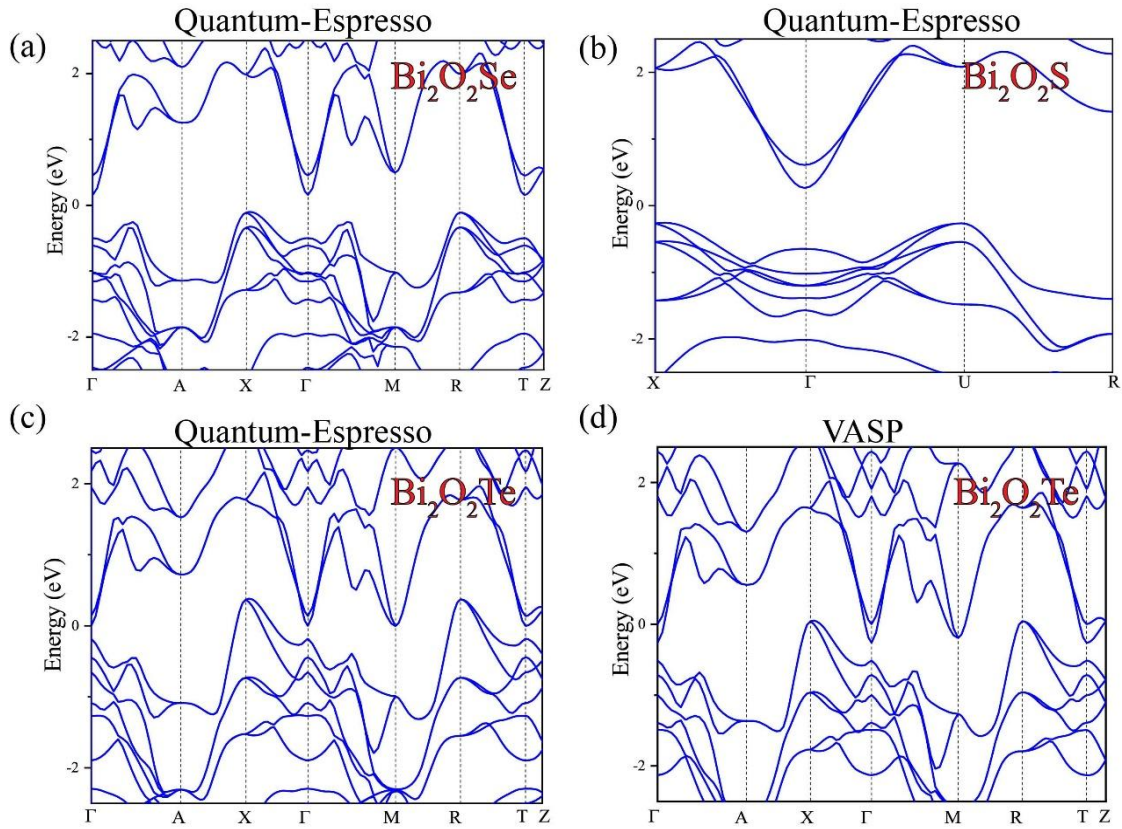

**Figure S2.** Electronic band structures of (a) $Bi_2O_2Se$, (b) $Bi_2O_2S$, (c) $Bi_2O_2Te$ calculated by LDA method using the Quantum Espresso package, and (d) $Bi_2O_2Te$ calculated by LDA method using the VASP. SOC effect was considered here.

## 3. The Calculated atomic displacements for all remaining Raman-active modes in $Bi_2O_2S$ ($D_{2h}$ symmetry).

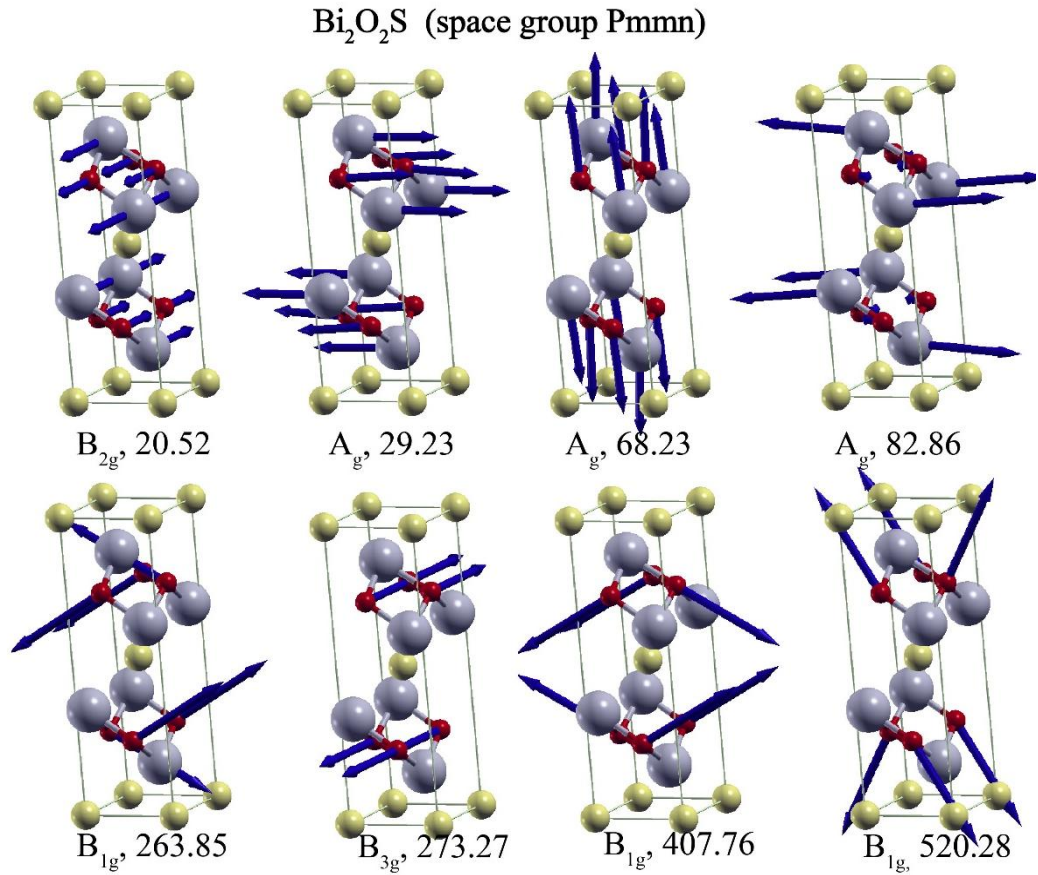

**Figure S3.** The calculated polarization vectors of all Raman-active modes in $Bi_2O_2S$ except the modes which have been shown in the main paper.

## 4. Separate the $A_{1g}$ and $B_{1g}$ mode in theory

Considering the light polarization configuration, $e_i$ and $e_s$ are the polarization vectors of the incoming and scattered photons, respectively. Selection rules help us to know that the intensity of the $A_{1g}$ mode is independent of the sample orientation, but the $B_{1g}$ mode is strongly related ($I_{B1g}(\theta) \sim c^2 \sin^2(\theta+2\beta)$, where $\theta = \angle(e_i, e_s)$, $\beta = \angle(e_i, x)$). The diagram illustration was shown in Figure S3 (a). Figure S3 (b) gives the intensity of the $A_{1g}$ and $B_{1g}$ modes as a function of the crystal orientation with respect to the laboratory axes $x_0$ and $y_0$. It can be seen that, in the parallel polarization configuration ($\theta=0°$), the intensity of $B_{1g}$ mode is maximal when the orientation of the sample is parallel to $e_i$ ($\beta=0°$), then gradually decreases to zero for $\beta=45°$. But the $A_{1g}$ always keeps constant when the orientation of the sample changes. In the crossed polarization configuration ($\theta=90°$), the intensity of $B_{1g}$ mode gradually increases from zero ($\beta=0°$) to its maximal value for $\beta=45°$, whereas $A_{1g}$ mode vanishes in this configuration. So we could choose specified sample orientation ($\beta=45°$, $x'=1/\sqrt{2}[110]$, $y'=1/\sqrt{2}[1-10]$) and various polarized scattering configuration to assign four different modes in the $Bi_2O_2Se$ ($Bi_2O_2Te$) system (see Figure S3 (c)). This hypothesis could be verified by future polarized Raman scattering experiments when obtain a good sample.

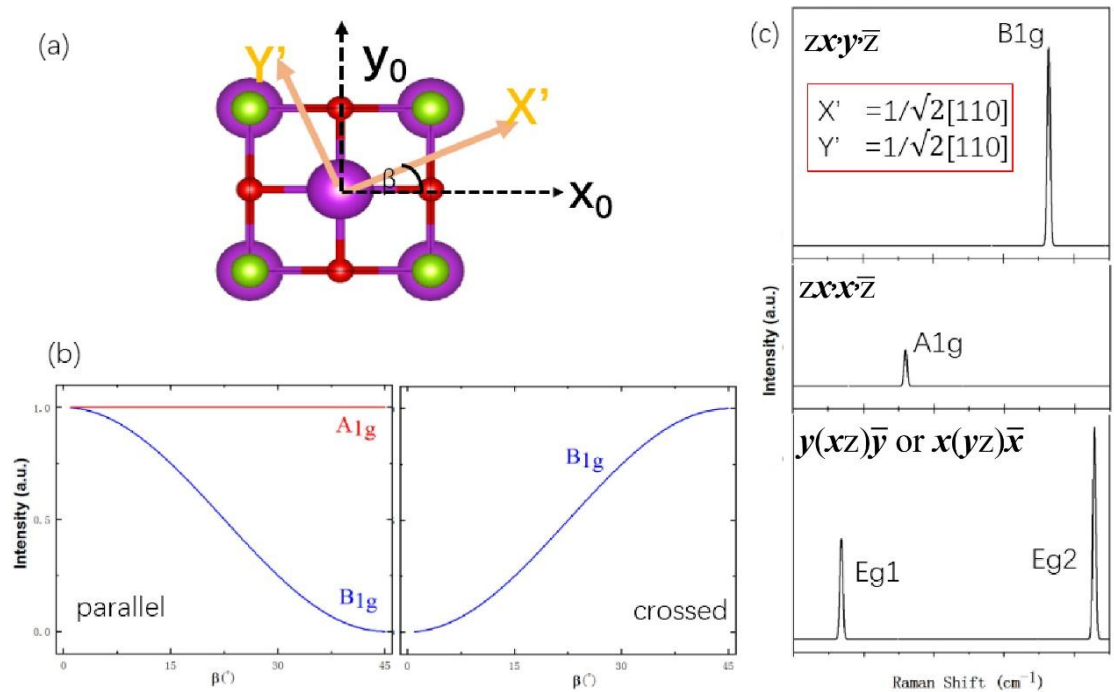

**Figure S4.** (a) The Scheme for considering the sample orientation $\beta$ with respect to the laboratory axes $x_0$ and $y_0$. (b) variation of the intensities I [$A_{1g}$] and I [$B_{1g}$] versus sample orientation $\beta$ in the parallel and crossed polarization configurations. (c) the predicted polarized Raman scattering

spectra of Bi2O2Se (Bi2O2Te) in various scattering configurations ($x$ = [110], $y$ = [010], $x'=1/\sqrt{2}$[110], $y'=1/\sqrt{2}$[1-10], $z$ = [001]).

## 5. Detailed process for deducing the frequency response functions with small strains

For degenerate modes, it will split into two modes under uniaxial strain. Therefore, we could suppose the Hamiltonian with the energies of two modes as the eigenvalues has such form:

$$H = \begin{bmatrix} a & c \\ c & b \end{bmatrix} \quad (1)$$

where $a$, $b$ and $c$ are matrix elements. The solutions of Eq.(1) are:

$$\omega_{1,2} = \frac{(a+b) \pm \sqrt{(a-b)^2 + 4c^2}}{2} \quad (2)$$

The two frequency modes degeneracy conditions are: $a=b$ and $c=0$. Under small strains, we could suppose that the parameter $a$, $b$ and $c$ expands like:

$$\begin{cases} a = a_0 + \alpha_1 \varepsilon_{xx} + \beta_1 \varepsilon_{yy} + \chi_1 \gamma \\ b = a_0 + \alpha_2 \varepsilon_{xx} + \beta_2 \varepsilon_{yy} + \chi_2 \gamma \\ c = c_0 + \alpha_3 \varepsilon_{xx} + \beta_3 \varepsilon_{yy} + \chi_3 \gamma \end{cases} \quad (3)$$

To substitute Eq. (3) into Eq. (2), we could have:

$$\omega_{1,2} = a_0 + \frac{(\alpha_1 + \alpha_2)\varepsilon_{xx} + (\beta_1 + \beta_2)\varepsilon_{yy} + (\chi_1 + \chi_2)\gamma}{2}$$
$$\pm \frac{\sqrt{[(\alpha_1 - \alpha_2)\varepsilon_{xx} + (\beta_1 - \beta_2)\varepsilon_{yy} + (\chi_1 - \chi_2)\gamma]^2 + 4(\alpha_3 \varepsilon_{xx} + \beta_3 \varepsilon_{yy} + \chi_3 \gamma)^2}}{2} \quad (4)$$

Then we could simplify Eq. (4) by symmetry restrictions of the $D_{4h}$ point group:

(1) Inversion operation: to change the sign of shear strain ($\gamma \rightarrow -\gamma$), the symmetry is not broken.

That is:

$$\omega_{1,2} = a_0 + \frac{(\alpha_1+\alpha_2)\varepsilon_{xx}+(\beta_1+\beta_2)\varepsilon_{yy}+(\chi_1+\chi_2)\gamma}{2}$$

$$\pm \frac{\sqrt{[(\alpha_1-\alpha_2)\varepsilon_{xx}+(\beta_1-\beta_2)\varepsilon_{yy}+(\chi_1-\chi_2)\gamma]^2+4(\alpha_3\varepsilon_{xx}+\beta_3\varepsilon_{yy}+\chi_3\gamma)^2}}{2}$$

$$\equiv a_0 + \frac{(\alpha_1+\alpha_2)\varepsilon_{xx}+(\beta_1+\beta_2)\varepsilon_{yy}-(\chi_1+\chi_2)\gamma}{2}$$

$$\pm \frac{\sqrt{[(\alpha_1-\alpha_2)\varepsilon_{xx}+(\beta_1-\beta_2)\varepsilon_{yy}-(\chi_1-\chi_2)\gamma]^2+4(\alpha_3\varepsilon_{xx}+\beta_3\varepsilon_{yy}-\chi_3\gamma)^2}}{2} \quad (5)$$

Thus we have

$$\begin{cases} \chi_1 + \chi_2 = 0 \\ (\alpha_1-\alpha_2)(\chi_1-\chi_2)+4\alpha_3\chi_3 = 0 \\ (\beta_1-\beta_2)(\chi_1-\chi_2)+4\beta_3\chi_3 = 0 \end{cases} \quad (6)$$

(2) Isotropic system: to apply the biaxial strain ($\varepsilon_{xx}=\varepsilon_{yy}=\varepsilon$, $\gamma=0$), the symmetry is conserved and no split happens. That is:

$$\frac{\sqrt{[(\alpha_1-\alpha_2)\varepsilon+(\beta_1-\beta_2)\varepsilon]^2+4(\alpha_3\varepsilon+\beta_3\varepsilon)^2}}{2} \equiv 0 \quad (7)$$

Thus, we have

$$\begin{cases} \alpha_1 + \beta_1 = \alpha_2 + \beta_2 \\ \alpha_3 + \beta_3 = 0 \end{cases} \quad (8)$$

(3) The system is conserved after a rotation of 90°: At this time, the strain after rotation will be:

$$\begin{bmatrix} \varepsilon_{xx}' & \gamma' \\ 0 & \varepsilon_{yy}' \end{bmatrix} = \begin{bmatrix} \cos\theta & \sin\theta \\ -\sin\theta & \cos\theta \end{bmatrix} \begin{bmatrix} \varepsilon_x' & \gamma' \\ 0 & \varepsilon_y' \end{bmatrix} \begin{bmatrix} \cos\theta & -\sin\theta \\ \sin\theta & \cos\theta \end{bmatrix}$$

$$= \begin{bmatrix} \varepsilon_{yy} & 0 \\ -\gamma & \varepsilon_{xx} \end{bmatrix} \quad (9)$$

Then,

$$\omega_{1,2} = a_0 + \frac{(\alpha_1+\alpha_2)\varepsilon_{xx}+(\beta_1+\beta_2)\varepsilon_{yy}+(\chi_1+\chi_2)\gamma}{2}$$

$$\pm \frac{\sqrt{[(\alpha_1-\alpha_2)\varepsilon_{xx}+(\beta_1-\beta_2)\varepsilon_{yy}+(\chi_1-\chi_2)\gamma]^2+4(\alpha_3\varepsilon_{xx}+\beta_3\varepsilon_{yy}+\chi_3\gamma)^2}}{2}$$

$$\equiv a_0 + \frac{(\alpha_1+\alpha_2)\varepsilon_{yy}+(\beta_1+\beta_2)\varepsilon_{xx}-(\chi_1+\chi_2)\gamma}{2}$$

$$\pm \frac{\sqrt{[(\alpha_1-\alpha_2)\varepsilon_{yy}+(\beta_1-\beta_2)\varepsilon_{xx}-(\chi_1-\chi_2)\gamma]^2+4(\alpha_3\varepsilon_{yy}+\beta_3\varepsilon_{xx}-\chi_3\gamma)^2}}{2}$$

Based on Eq. (8), we have:

$$\alpha_1+\alpha_2 = \beta_1+\beta_2 \tag{10}$$

In all, we substitute Eq. (6), (8) and (10) into the initial Eq. (4), the frequencies of degenerate mode under small strain will be:

$$\begin{aligned}\omega_{1,2} &= \omega_0 + \frac{(\alpha_1+\alpha_2)}{2}(\varepsilon_{xx}+\varepsilon_{yy}) \\ &\pm \frac{\sqrt{\left[(\alpha_1-\alpha_2)(\varepsilon_{xx}-\varepsilon_{yy})+2\chi_1\gamma\right]^2+4[\alpha_3(\varepsilon_{xx}-\varepsilon_{yy})+\chi_3\gamma]^2}}{2} \\ &= \omega_0 + \frac{(\alpha_1+\alpha_2)}{2}(\varepsilon_{xx}+\varepsilon_{yy}) \\ &\pm \frac{\sqrt{[(\alpha_1-\alpha_2)^2+4\alpha_3^2](\varepsilon_{xx}-\varepsilon_{yy})^2+4(\chi_1^2+\chi_3^2)\gamma^2}}{2}\end{aligned} \tag{11}$$

Define $\bar{k} = \frac{\alpha_1+\alpha_2}{2}, \Delta k = \sqrt{\frac{1}{4}(\alpha_1-\alpha_2)^2+\alpha_3^2}; k_\gamma = \sqrt{\chi_1^2+\chi_3^2}$, we could further simplify Eq. (11) as:

$$\omega_{1,2}(\varepsilon_{xx},\varepsilon_{yy},\gamma) = \omega_0 + \bar{k}(\varepsilon_{xx}+\varepsilon_{yy}) \pm \sqrt{[\Delta k(\varepsilon_{xx}-\varepsilon_{yy})]^2+(k_\gamma\gamma)^2} \tag{12}$$

This gives a universal formula between small strains and frequencies of the split doubly degenerate modes for $D_{4h}$ symmetry systems.

# 6. Calculated 2D elastic constant evaluation for monolayer Bi$_2$O$_2$Se and Bi$_2$O$_2$Te.

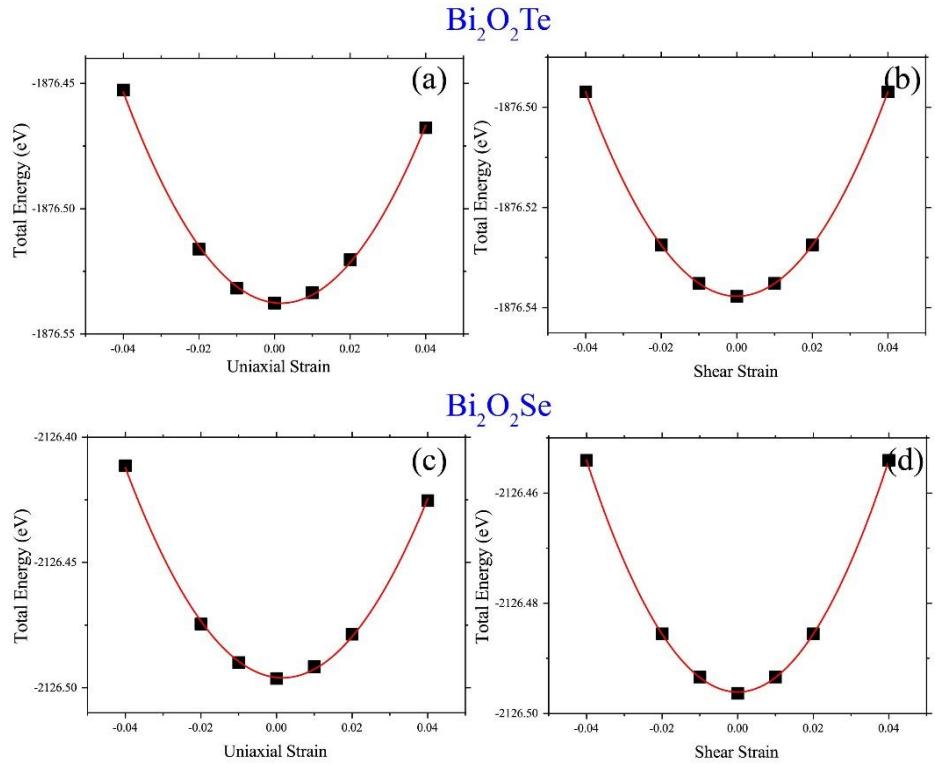

**Figure S5.** 2D elastic constant evaluation (Total energy with respect to the uniaxial strain and shear strain) for the (a, b) Bi$_2$O$_2$Te and (c, d) Bi$_2$O$_2$Se. The quadratic fit gives the 2D elastic constant.

7. **The remaining evolutions of all Raman modes with small strains (both uniaxial and shear strains) for bulk and monolayer $Bi_2O_2Se$ ($Bi_2O_2Te$).**

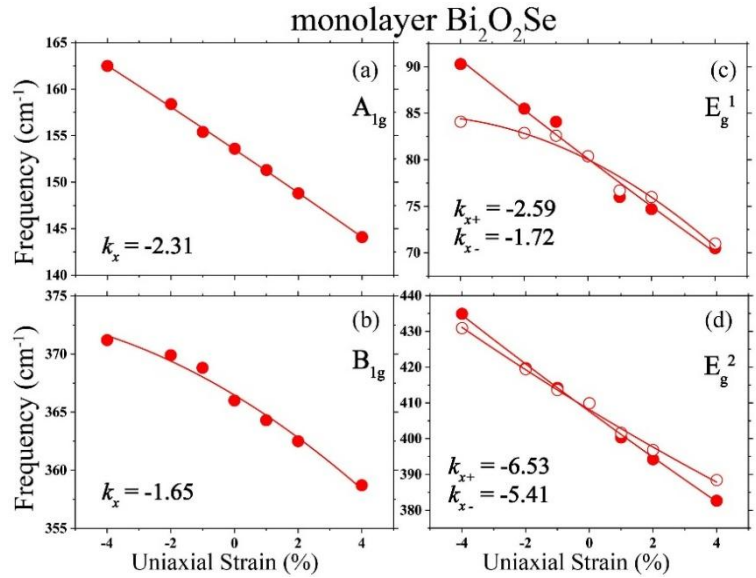

**Figure S6.** The evolutions of Raman shift with uniaxial strain for monolayer $Bi_2O_2Se$.

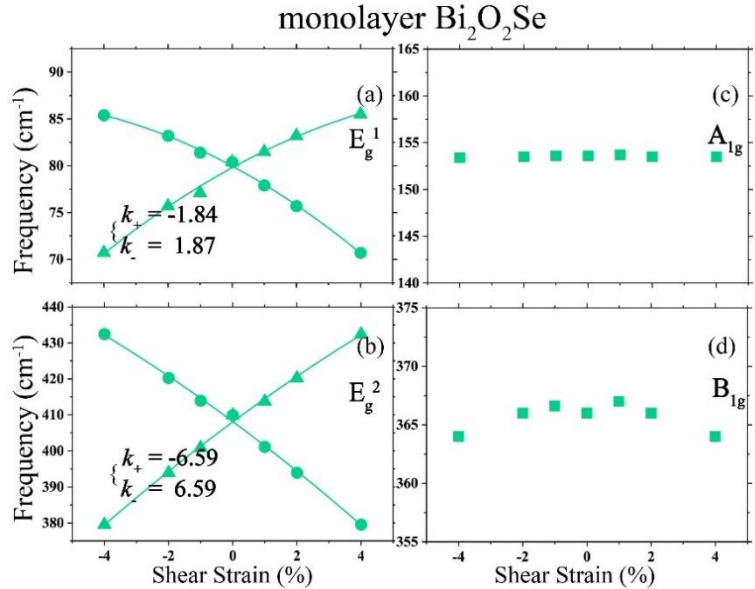

**Figure S7.** The evolutions of Raman shift with shear strain for monolayer $Bi_2O_2Se$.

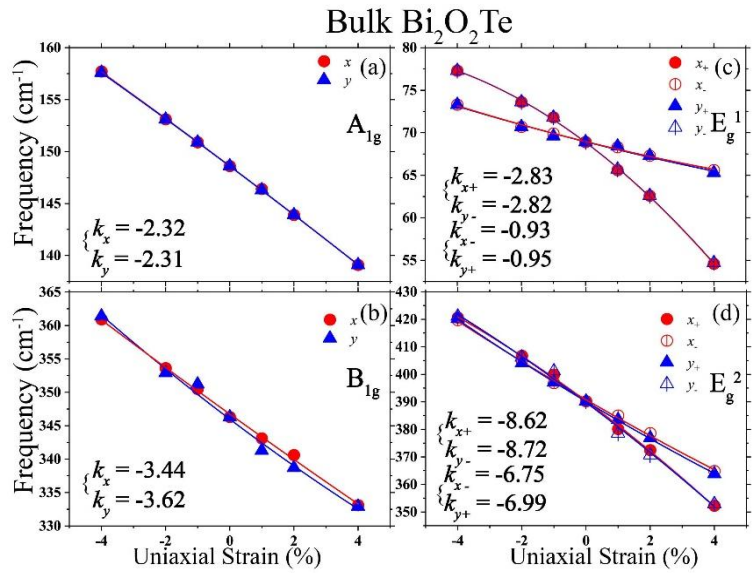

**Figure S8.** The evolutions of Raman shift with uniaxial strain for bulk $Bi_2O_2Te$.

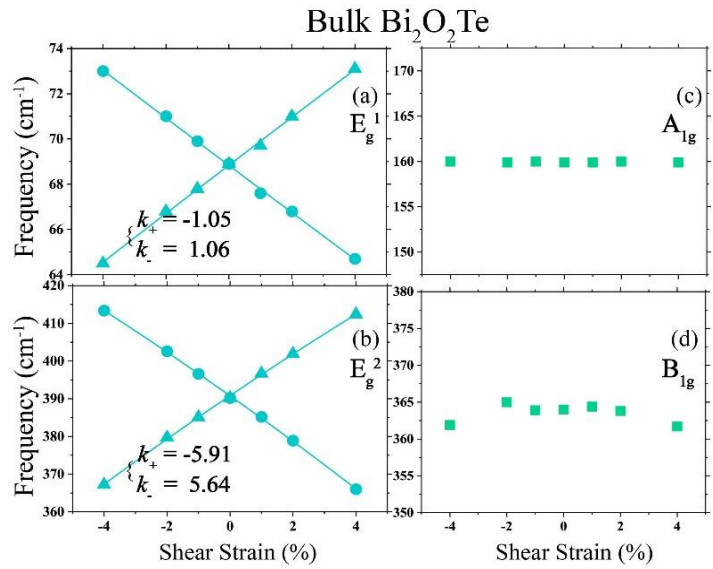

**Figure S9.** The evolutions of Raman shift with shear strain for bulk $Bi_2O_2Te$.

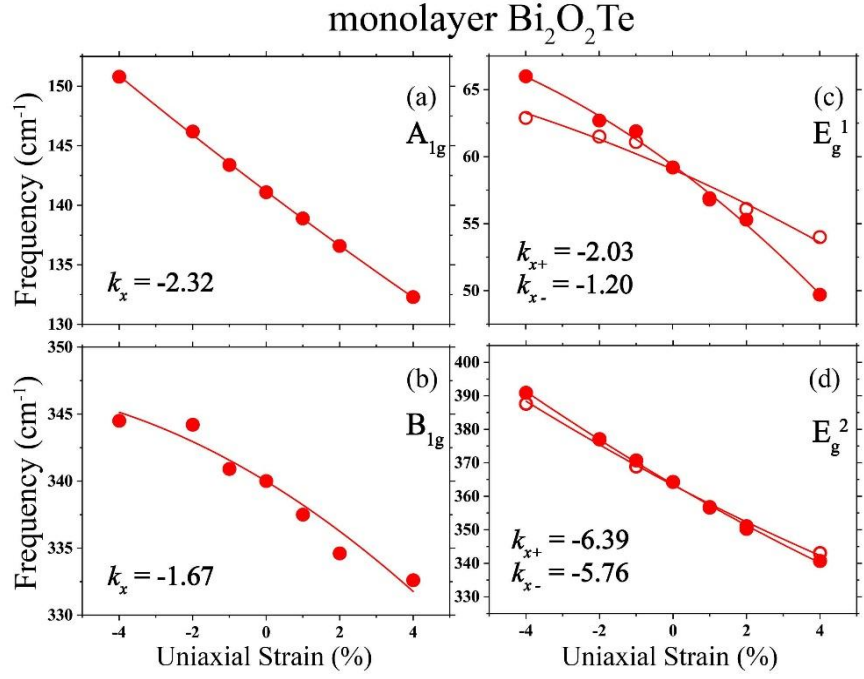

**Figure S10.** The evolutions of Raman shift with uniaxial strain for monolayer $Bi_2O_2Te$.

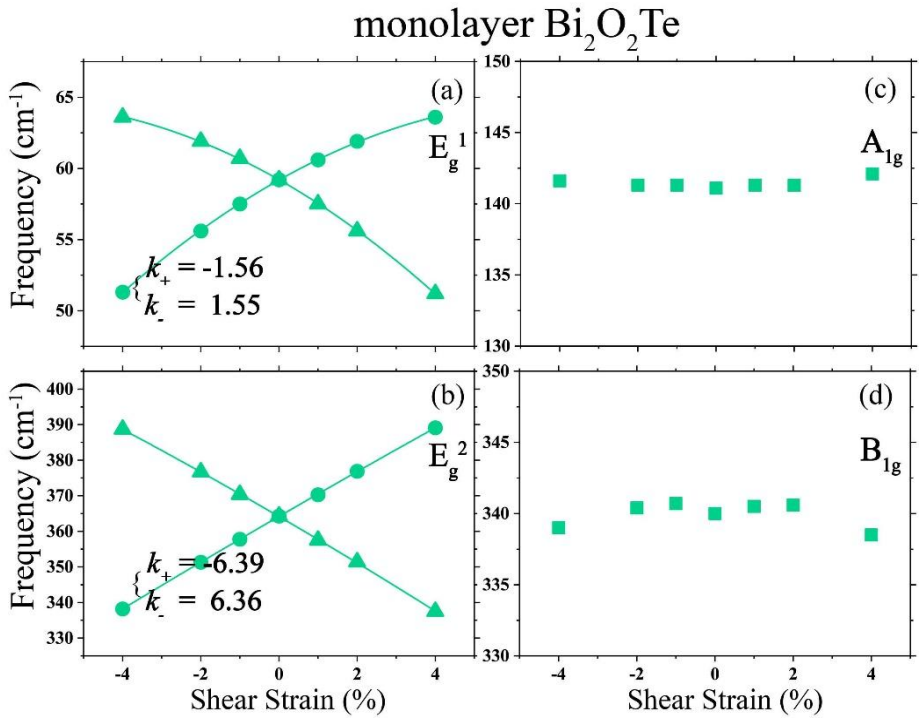

**Figure S11.** The evolutions of Raman shift with shear strain for monolayer $Bi_2O_2Te$.

**Table S1. Determined slopes for four Raman-active modes in bulk and monolayer $Bi_2O_2Se$ and $Bi_2O_2Te$ under applied strain $\varepsilon_{xx}$ and $\gamma$.**

| Slope $k$ | $Bi_2O_2Se$ | | | | $Bi_2O_2Te$ | | | |
|---|---|---|---|---|---|---|---|---|
| | Bulk | | Monolayer | | Bulk | | Monolayer | |
| | $k_{xx}$ | $k_\gamma$ | $k_{xx}$ | $k_\gamma$ | $k_{xx}$ | $k_\gamma$ | $k_{xx}$ | $k_\gamma$ |
| $E_g^1$ | −2.37; −0.66 | −0.81; 0.81 | −2.59; −1.72 | −1.84; 1.87 | −2.83; −0.93 | −1.05; 1.06 | −2.03; −1.20 | -1.56; 1.55 |
| $A_{1g}$ | −2.32 | 0 | −2.31 | 0 | −2.32 | 0 | −2.32 | 0 |
| $B_{1g}$ | −3.00 | 0 | −1.65 | 0 | −3.44 | 0 | −1.67 | 0 |
| $E_g^2$ | −8.47; −6.97 | −6.36; 6.29 | −6.53; -5.41 | −6.59; 6.59 | −8.62; −6.75 | −5.91; 5.64 | −6.39; −5.76 | -6.39; 6.36 |

8. **The corresponding *θ*-dependent frequency change of the degenerate modes in monolayer and bulk Bi$_2$O$_2$Se (Bi$_2$O$_2$Te).**

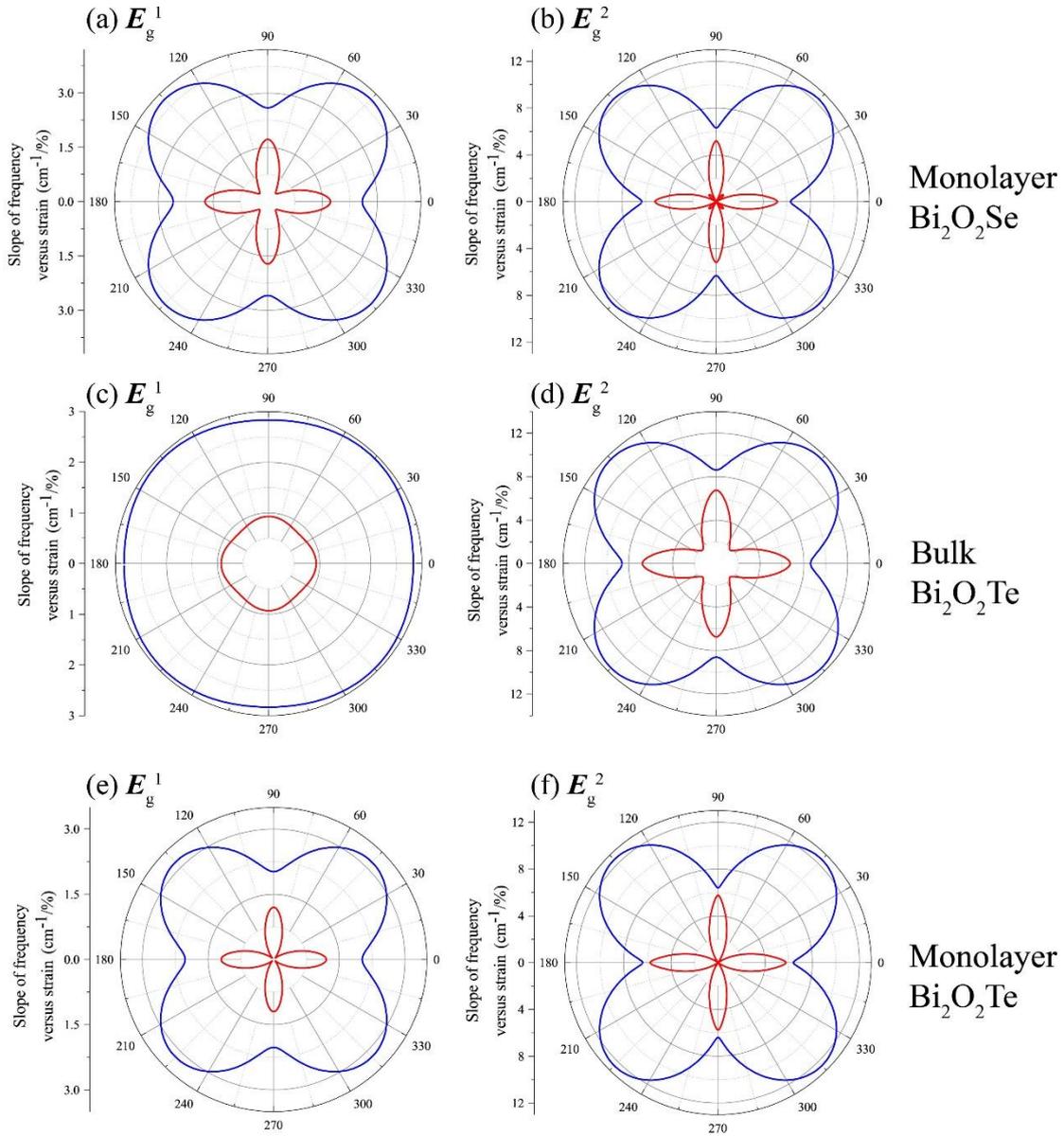

**Figure S12.** The corresponding *θ*-dependent frequency change of degenerate modes in monolayer and bulk Bi$_2$O$_2$Se and Bi$_2$O$_2$Te.